\documentstyle[aps,fleqn]{revtex}
\begin{document} 
\draft

%% Include following two lines for Journal Style

%\twocolumn[\hsize\textwidth\columnwidth\hsize  %**Journal
%\csname @twocolumnfalse\endcsname              %**Journal

\title{Effects of Imperfect Gate Operations  
 in Shor's Prime Factorization Algorithm} 
\author{Hao Guo$^{1,2}$, Gui-Lu Long$^{1,2,3,4,5}$ and Yang Sun$^{1,2,6,7}$} 
\address{  
$^1$Department of Physics, Tsinghua University,  
 Beijing 100084\\ 
 $^2$Key Laboratory for Quantum Information and Measurements,  MOE \\ 
$^3$Institute of Theoretical Physics, Chinese Academy of Sciences,  
 Beijing 100080, P.R. China\\ 
$^4$Centre for Nuclear Theory, Lanzhou National Laboratory of Heavy Ions\\  
 Chinese Academy of Sciences, Lanzhou 740000, P.R. China\\  
$^5$ Center of Atomic, Molecular and Nanosciences, Tsinghua University, Beijing 100084\\
$^6$Department of Physics, Xuzhou Normal University,
 Xuzhou, Jiangsu 221009
$^7$Department of Physics and Astronomy, University of Tennessee, 
 Knoxville, TN 37996, U.S.A.
} 

\date{\today}
\maketitle

\begin{abstract} 
The effects of imperfect gate operations 
in implementation of Shor's prime factorization algorithm are investigated.  
The gate imperfections may be 
classified into three categories:  
the systematic error, the random error, and the one with combined  
errors. It is found that Shor's algorithm is robust against the 
systematic errors but is 
vulnerable to the random errors.   
Error threshold is given to the algorithm for a given number 
$N$ to be factorized. 
\end{abstract} 


%%  Include the following line for Journal style
%]  %**Journal
 
\section{Introduction} 
Shor's factorization algorithm \cite{Shor94} 
is a very important quantum algorithm, through which   
one has demonstrated the power of quantum computers. 
It has greatly promoted the worldwide research in quantum computing  
over the past few years. In practice, 
however, quantum systems are subject to 
influence of environment, and in addition, quantum gate operations  
are often imperfect \cite{EJ96,S2}. Environment influence on the system  
can cause decoherence of quantum states, and gate imperfection leads to  
errors in quantum computing. Thanks to Shor's another important work, 
in which he showed   
that quantum error correlation can be corrected \cite{S3}.  
With quantum error correction scheme, errors arising from both decoherence  
and imperfection can be corrected. 
 
There have been several works on the effects of decoherence 
on Shor's algorithm. 
Sun {\it et al.} discussed the effect of decoherence on the algorithm  
by modeling the environment \cite{S4}. 
Palma studied the effects of both decoherence  
and gate imperfection in ion trap quantum computers \cite{S5}. 
There have also been many other studies on the quantum algorithm  
\cite{S6,S7,S8,S9}. 

The error correction scheme uses available resources.  
Thus it is important to study the robustness of the algorithm itself  
so that one can strike a balance between the amount of quantum error  
correction and the amount of qubits available. 
In this paper, we investigate the effects of gate  
imperfection on the efficiency of Shor's factorization algorithm.  
The results may guide us in practice to suppress deliberately those errors 
that influence the algorithm most sensitively. 
For those errors that do not affect the algorithm very much, 
we may ignore them as a good approximation. 
In addition, study of the robustness of algorithm to errors is important  
where one can not apply the quantum error correction at all, for instance,  
in cases that there are not enough qubits available. 
 
The paper is organized as follows. Section II is devoted to 
an outline of Shor's algorithm  
and different error's modes. In Section III, we present the results.  
Finally, a short summary is given in Section IV.

\section{Shor's algorithm and error's Modes} 
Shor's algorithm consists of the following steps: 
 
1) preparing a superposition of evenly distributed states   
$$|\psi\rangle=\frac{1}{\sqrt{q}}\sum_{a=0}^{q-1}|a\rangle|0\rangle ,$$ 
where $q=2^L$ and $N^2\leq{q}\leq2N^2$ 
with $N$ being the number to be factorized; 
 
2) implementing $y^{a}$mod$N$ and putting the results into the 2nd register   
$$|\psi_{1}\rangle
=\frac{1}{\sqrt{q}}\sum_{a=0}^{q-1}|a\rangle|y^{a}{\rm mod}N\rangle ;$$ 
 
3) making a measument on the 2nd register; The state of the register is then  
$$|\phi_{2}\rangle=\frac{1}{\sqrt{A+1}}\sum_{j=0}^{A}|jr+l\rangle|z 
=y^{l}=y^{jr+l}mod N\rangle$$ 
where $j\leq\left [\frac{q-l}{r}\right ]=A$.
 
4) performing discrete Fourier transformation (DFT) on the first register  
$|\phi_{3}\rangle 
=\left(\sum_{c}\tilde f \left(c \right)|c\rangle\right)|z\rangle$,  
where 
$$\tilde f\left(c\right) 
=\frac{\sqrt{r}}{q} 
\sum_{j=0}^{\frac{q}{r}-1}exp\left (\frac{2\pi{i(jr+l)}}{q}\right ) 
=\frac{\sqrt{r}}{q}e^{\frac{2\pi{ilc}}{q}} 
\sum_{j=0}^{\frac{q}{r}-1}exp\left (\frac{2\pi{ijrc}}{q}\right ) .$$ 
This term is nonzero only when $c=k\frac{q}{r}$, with $k=0,1,2...r-1$,  
which correspond to the peaks of the distribution in the measured results,  
and thus this term becomes  
$\tilde f(c)=\frac{1}{\sqrt{r}}e^{\frac{2\pi{ilc}}{q}}$. 
The Fourier transformation is important because it makes the state  
in the first register the same for all possible values in the 2nd register. 
The DFT is constructed by two basic gate  
operations: the single bit gate operation  
$A_j=\frac{1}{\sqrt{2}}\left(\begin{array}{cc}1 & 1\\1 &-1 \end{array}\right ) $, 
which is also called the Walsh-Hadmard transformation, 
and the 2-bits controlled rotation 
\begin{displaymath} 
B_{jk}=\left (\begin{array}{cccc} 
                   1 & 0 & 0 & 0\\ 
                   0 & 1 & 0 & 0\\ 
                   0 & 0 & 1 & 0\\ 
                   0 & 0 & 0 & e^{i\theta_{jk}} 
                \end{array} \right ) 
\end{displaymath}   
with $\theta_{jk}=\frac{\pi}{2^{k-j}}$. 
The gate sequence for implementing DFT is  
$$(A_{q-1})(B_{q-2q-1}A_{q-2}) \ldots (B_{0q-1}B_{0q-2} \ldots B_{01}A_{0}). $$ 

Errors can occur in both $A_j$ and $B_{jk}$. $A_j$ is actually a rotation  
about y-axis through $\frac{\pi}{2}$  
$$A_{j}(\theta)=e^{\frac{i}{\hbar}S_{y}\theta} 
=I\cos(\frac{\theta}{2})-i\sin(\frac{\theta}{2})\sigma_{y} 
=\left (\begin{array}{cc}  
              \cos(\frac{\theta}{2}) &   -\sin(\frac{\theta}{2})\\                          %%@
\sin(\frac{\theta}{2}) &   \cos(\frac{\theta}{2}) 
              \end{array}\right ) .$$ 
If the gate operation is not perfect,  
the rotation is not exactly $\frac{\pi}{2}$. 
In this case, $A_j$ is a rotation of $\frac{\pi}{2}+2\delta$ 
$$A_{j}(\delta)=\frac{1}{\sqrt{2}}\left (\begin{array}{cc}  
              \cos(\delta)-\sin(\delta) & -(\sin(\delta)+\cos(\delta))\\ 
              \sin(\delta)+\cos(\delta) & \cos(\delta)-\sin(\delta) 
              \end{array}\right ) .$$ 
If $\delta$ is very small, we have: 
$$A_{j}(\theta)=\frac{1}{\sqrt{2}}\left (\begin{array}{cc}  
              1-\delta & -(1+\delta)\\ 
              1+\delta+ & 1-\delta 
              \end{array}\right ) .$$ 
Similarly, errors in $B_{jk}$ can be written as  
\begin{displaymath}  
B_{jk}=\left (\begin{array}{cccc} 
                   1 & 0 & 0 & 0\\ 
                   0 & 1 & 0 & 0\\ 
                   0 & 0 & 1 & 0\\ 
                   0 & 0 & 0 & e^{i(\theta_{jk}+\delta)} 
                \end{array} 
\right ) . 
\end{displaymath}  
With these errors, the DFT becomes 
\begin{eqnarray} 
|a\rangle\rightarrow\frac{1}{\sqrt{q}} 
\sum_{c=0}^{q-1}e^{i(\frac{2\pi}{q/c}+\delta_{c})a}
(1+\delta_{c}')|\tilde c\rangle 
=\frac{1}{\sqrt{q}}\sum_{c=0}^{q-1}e^{i(\frac{2{\pi}c}{q} 
+\delta_{c})a}(1+\delta_{c}')|\tilde c\rangle , 
\end{eqnarray} 
where $\delta_{c}$ and $\delta_{c}'$  
denote the error of $A_{j}$ and $B_{jk}$, respectively. 

Let us assume the following error modes:  
1) systematic errors, where $\delta_{c}$ or $\delta_{c}'$ in (1) can only have  
systematic errors (EM$_{1}$); 2) random errors (EM$_{2}$), for which   
we assume that $\delta_{c}$ or $\delta_{c}'$ can only be random errors  
of the Gaussian or the uniform type;  
3) coexistence of both systematic and random errors (EM$_{3}$).  
In the next section, we shall present the results of numerical simulations  
and discuss the effects of imperfect gate operation on the DFT algorithm, 
and thus on the Shor's algorithm. 
 
\section{Influence of Imperfect Gate Operations} 
 
We first discuss the influence of imperfect gate operations 
in the initial preparation 
\begin{displaymath} 
\begin{array}{ll} 
A_{l-1}A_{l-2}...A_{0}|0...0\rangle 
&=\frac{1}{\sqrt{2}}(|0\rangle+|1\rangle+\delta_{1}(|0\rangle-|1\rangle)) 
\otimes(|0\rangle+|1\rangle+\delta_{2}(|0\rangle-|1\rangle))\otimes \ldots 
\otimes(|0\rangle+|1\rangle+\delta_{n}(|0\rangle-|1\rangle))\\ 
&\doteq\frac{1}{\sqrt{2^{l}}} 
\sum_{i_{1}i_{2}...i_{n} =0}^{1}|i_{1}i_{2}...i_{n}\rangle  
+\frac{1}{\sqrt{2^{l}}}\sum_{R=1}^{n}\delta_{n} 
\sum_{i_{1}i_{2}...i_{n}=0}^{1}  
(|i_{1}..i_{R-1}0i_{R+1}..i_{n}\rangle-|i_{1}..i_{R-1}1i_{R+1}..i_{n}\rangle 
\end{array} 
\end{displaymath} 
If the errors are systematic, for instance,  
caused by the inaccurate calibration of the rotations, 
then $\delta_{1}=\delta_{2}= \ldots =\delta_{n}=\delta$. 
In this case, we can write the 2nd term as 
$$|\psi\rangle=\frac{1}{\sqrt{2^{l}}}\delta 
\sum_{i_{1}i_{2}...i_{n}=0}^{1}(2s-n)|i_{1}i_{2}...i_{n}\rangle ,$$ 
where $s$ stands for the number of 1's, and $2s-n=s-(n-s)$ is the difference  
in the number of 1's and 0's.  
Thus the results after the first procedure is 
\begin{eqnarray} 
\frac{1}{\sqrt{2^{l}}}\sum_{a=0}^{2^{L}-1}(|a\rangle+\delta(2s-n)|a\rangle) 
=\frac{1}{\sqrt{2^{l}}}\sum_{a=0}(1+\delta_{a})|a\rangle .     
\end{eqnarray} 
This implies that after the procedure, 
the amplitude of each state is no longer equal,  
but have slight difference. Combining the effect in the initialization and in  
the DFT, we have  
$$(1+\delta_{a})(1+\delta_{c})e^{i(\frac{2{\pi}c}{q}+\delta_{c}')a} 
\doteq(1+\delta'')e^{i(\frac{2{\pi}c}{q}+\delta_{c}')a} ,$$    
where $ \delta_{c}''=\delta_{c}+\delta_{a}$. 
In the DFT, we have 
$$|\psi\rangle\Rightarrow\frac{\sqrt{r}}{q}\sum_{c=0}^{q-1} 
\sum_{j=0}^{\frac{q}{r}-1}(1+\delta_{j}) 
e^{i(\frac{2{\pi}c}{q}+\delta_{j}')(jr+l)}|\tilde c\rangle ,$$
where we have rewrite $\delta''$ as $\delta_j$ here. 
Let $P_{c}$ denote the probability of getting the state $|\tilde c\rangle$  
after we perform a measurement, we have 
\begin{eqnarray} 
P_{c}&=&\frac{r}{q^{2}}\sum_{m=0}^{\frac{q}{r}-1} 
\sum_{k=0}^{\frac{q}{r}-1}(1+\delta_{m})(1+\delta_{k}) 
e^{i(\frac{2{\pi}c}{q}+\delta_{m}')(mr+l)}{\times} 
e^{-i(\frac{2{\pi}c}{q}+\delta_{k}')(kr 
+l)} \nonumber\\ 
&=&\frac{r}{q^{2}}\sum_{m}\sum_{k}(1+\delta_{m})(1+\delta_{k}) 
\cos[\frac{2{\pi}c}{q}r(m-k)+(mr+l)\delta_{m}'-(kr+l)\delta_{k}']         
\end{eqnarray} 
From Eq. (3), we find that after the last measurement,  
each state can be extracted with a probability which is nonzero,  
and the offset $l$ can't be eliminated. 
 
Eq. (3) is very complicated, so we will make some predigestions 
to discuss different error modes for convenience.  
Generally speaking, the influence of exponential error $\delta_{j}$  
is more remarkable than $\delta_{j}$,  
so we can omit the error $\delta_{j}$, thus 
\begin{center} 
DFT$_{q}$ $|\phi\rangle=\frac{\sqrt{r}}{q} 
\sum_{c=0}^{q-1}\sum_{j=0}^{\frac{q}{r}} 
e^{i(\frac{2{\pi}c}{q}+\delta_{j}')(jr+l)}|c\rangle$ .
\end{center} 

\subsection{Case 1} 
 
If only systematic errors (EM$_{1}$) are considered,  
namely, all the $\delta_{j}$'s are equal,  
then 
$\tilde f(c)$ can be given analytically 
\begin{eqnarray}
\tilde f(c)&=&\frac{\sqrt{r}}{q}\sum_{j=0}^{\frac{q}{r}-1} 
e^{i(\frac{2{\pi}c}{q}+\delta)(jr+l)} 
\nonumber\\ 
&=&\frac{\sqrt{r}}{q}e^{il(\frac{2{\pi}c}{q}+\delta)} 
\frac{1-e^{i(\frac{2{\pi}c}{q}+\delta)q}} 
{1-e^{i(\frac{2{\pi}c}{q}+\delta)r}}  
\label{fc}
\end{eqnarray}
The relative probability of finding $c$ is   
\begin{center}
P$_{c}=\left|   \tilde f(c)  \right|  ^{2}=\frac{r}{q^{2}} 
\frac{\sin^{2}(\frac{{\delta}q}{2})} 
{sin^{2}(\frac{{\pi}cr}{q}+\frac{{\delta}r}{2})}$, 
\end{center}
and if $c=k\frac{q}{r},$  
then 
\begin{center}
$P_{c}=\frac{r\sin^{2}(\frac{{\delta}q}{2})} 
{q^{2}\sin^{2}(\frac{{\delta}r}{2})}. $ 
\end{center}
It can be easily seen that $\lim_{\delta{\rightarrow}0}P_{c}=\frac{1}{r}$,  
which is just the case that no error is considered. 

When $\delta$ takes certain values, say,  
$\delta=\frac{2}{r}(k-\frac{r}{q})\pi$  
where $k$ is an integer, then the summation in Eq. (\ref{fc}) 
is on longer valid. In our simulation,  
$\delta$ does not take these values.  
Here we consider the case where $q=2^{7}=128$ and $r=4$. For comparisons, we have drawn %%@
the relative probability for obtaining state $c$ in Fig.1. for this given example. We have %%@
found
the following results:\\ 
(i) When $\delta$ is small, the errors do hardly influence the final result,  
for instance when $c=k\frac{q}{r}$, then 
$$\lim_{\delta_{\rightarrow}0}P_{c} 
=\lim_{\delta_{\rightarrow}0} 
\frac{r\sin^{2}(\frac{{\delta}q}{2})}{q^{2}\sin^{2}(\frac{{\delta}r}{2})} 
=\frac{1}{r} .$$ 
The probability distribution is almost identical to those without errors.\\ 
(ii) Let us increase $\delta$ gradually, from Fig.2,  
we see that a gradual change in the probability distribution takes place. 
(Here, we again consider the relative probabilities)  
When $\delta$ is increased to certain values,  
the positions of peaks change greatly. For instance at $\delta=0.05$,  
there appears a peak at c=127, whereas it is P$_{c}=0$  
when no systematic errors are present. In general,  
the influence of systematic errors on the algorithm is  
a shift of the peak positions.  
This influences the final results directly. 

\subsection{Case 2} 
 
When both random errors and systematic errors are present, 
we add random errors to the simulation.  
To see the effect of different mode of random errors,  
we use two random number generators. One is the Gaussian mode and  
the other is the uniform mode. In this case, the error has the form  
$\delta=\delta_{0}+s$, where $\delta_{0}$ is the systematic error.  
s has a probability distribution with respect to c,  
depending on the uniform or the Gaussian distribution.
When $\delta_{0}=0$, we have only random errors  
which is our error mode 2. When $\delta_{0}\not=0$, 
we have error mode 3. For the uniform distribution,  
$s\sim \pm s_{max}\times u(0,1)$ where $u(0,1)$ 
is evenly distributed in [0,1].  
$s_{max}$ indicates the maximum deviation from $\delta_{0}$.  
For Gaussian distribution, $s\sim N(0,\sigma_{0})$.  
Through the figure, we see the following: \\ 
(1) When only random errors are present
($\delta_{0}=0$),  
the peak positions are not affected by these random errors. However, 
different random error modes cause similar results.  The results for uniform random error %%@
mode are shown in Fig.3.
For the uniform distribution error mode, with increasing $\delta_{max}$,  
the final probability distribution of the final results become irregular.  
In particular, when $\delta_{max}$ is very large,  
all the patterns are destroyed and is hardly recognizable.  
Many unexpected small peaks appear. 
For the Gaussian distribution error mode, as shown in Fig.4, the influence of the error  
is more serious. This is because in Gaussian distribution,  
there is no cut-off of errors. Large errors can occur although  
their probability is small. The influence of $\sigma_{0}$ on the final results  
is also sensitive, because it determines the shape of the distribution.  
When $\sigma_{0}$ increases, the final probability distribution  
becomes very messy. A small change in $\sigma_{0}$ can cause  
a big change in the final results.\\ 
(2) When $\delta_{0}\not=0$, which corresponds to error mode 3,  
the effect is seen as to shift the positions of the peaks in addition to  
the influences of the random errors. 
 
\section{Summary} 

To summarize, we have analyzed the errors in Shor's factorization algorithm.  
It has been seen that the effect of the systematic errors is to shift  
the positions of the peaks, whereas the random errors change the shape of the  
probability distribution. For systematic errors, the shape of the  
distribution of the final results is hardly destroyed, though displaced.  
We can still use the result with several trial guesses  
to obtain the right results because the peak positions 
are shifted only slightly.  
However, the random errors are detrimental to the algorithm and  
should be reduced as much as possible.  
It is different from the case with Grover's algorithm where systematic errors 
are disastrous while random errors are less harmful \cite{S8}.

\baselineskip = 14pt
\bibliographystyle{unsrt}

\noindent Figure Captions:

\noindent Fig.1. Relative probability for finding state $c$ in the absence of errors.

\noindent Fig.2. The same as Fig.1. with systematic errors. In sub-figures (1), (2), (3), %%@
(4), $\delta$ are 0.02, 0.03, 0.05 respectively. In sub-figure (4), the curve with solid %%@
circles(with higher peaks) is the result with $\delta=0.1$, and the one without solid %%@
circles(with lower peaks) denotes the result with $\delta=0.33$.

\noindent Fig.3. The same as Fig.1. with uniform random errors. In sub-figures (1), (2), %%@
(3), (4), $s_{max}$ are set to 0.01, 0.03, 0.05, 0.1  respectively.

\noindent Fig.4. The same as Fig.1. with Gaussian random errors and systematic errors. In %%@
sub-figures (1), (2), and (3) $\tau$ are set to 0.01, 0.03 and 0.05 respectively, and %%@
$\delta_{0}=0$(without systematic errors).  In sub-figure (4), both systematic and random %%@
Gaussian errors exist, where $\delta_{0}=0.33$, $\tau=0.02$.\\
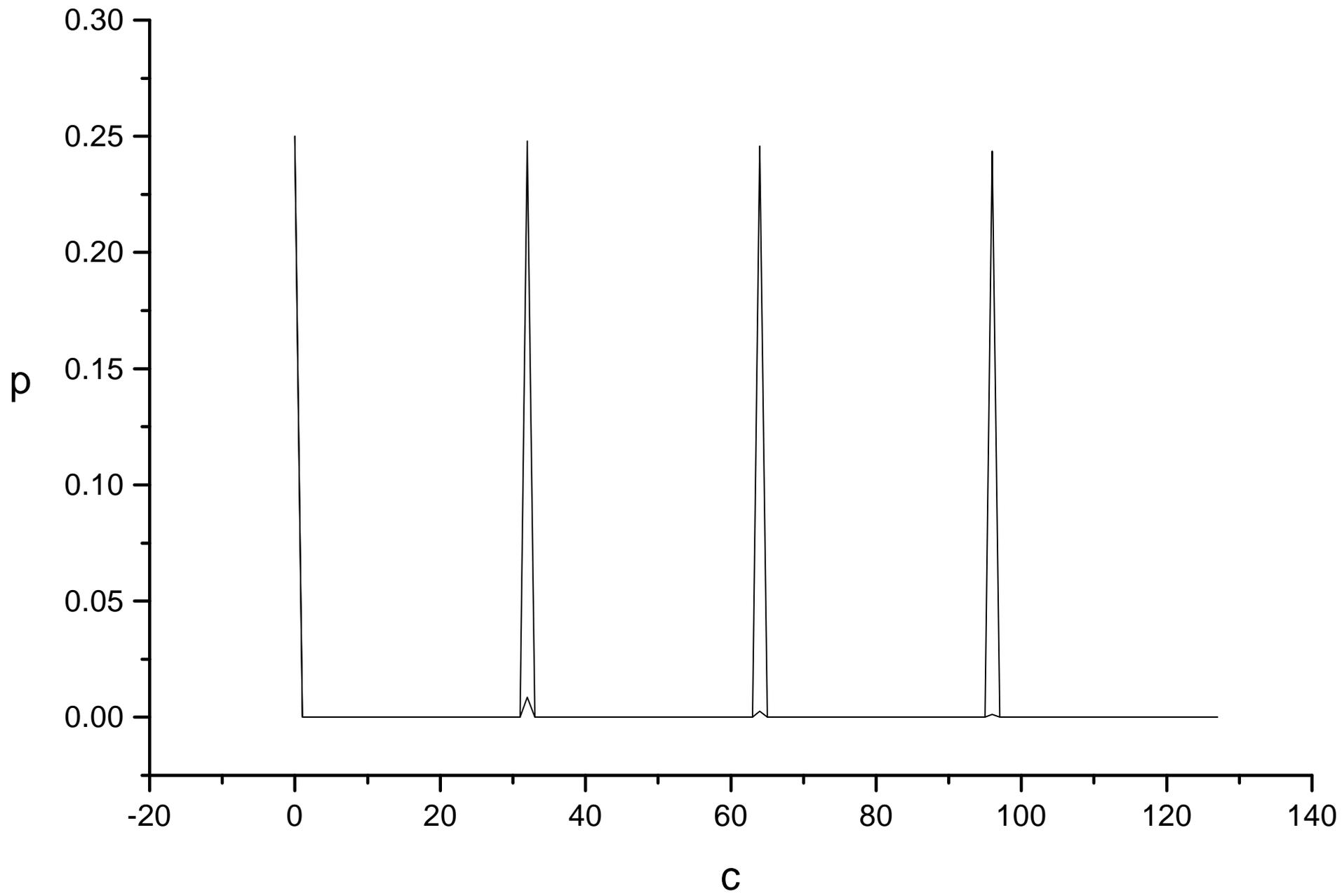
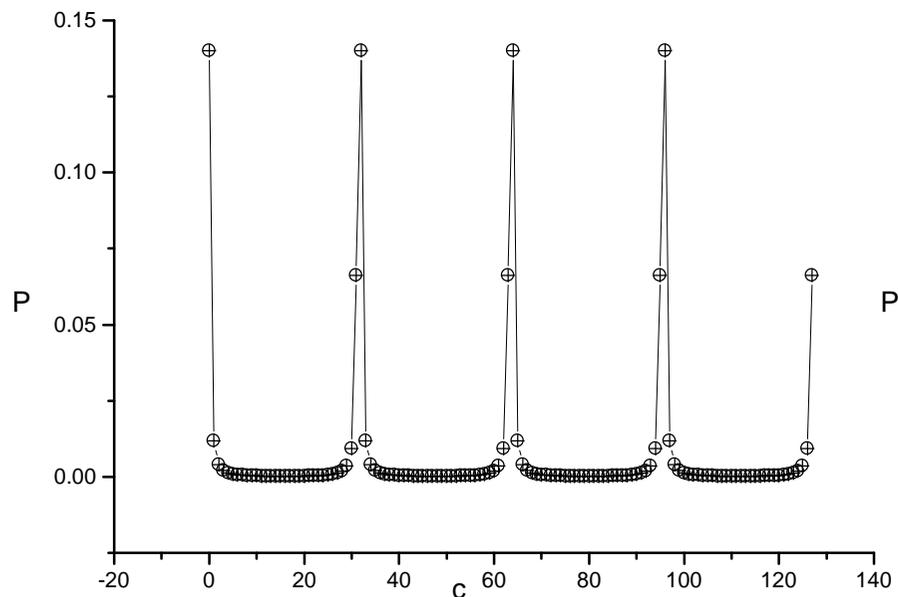
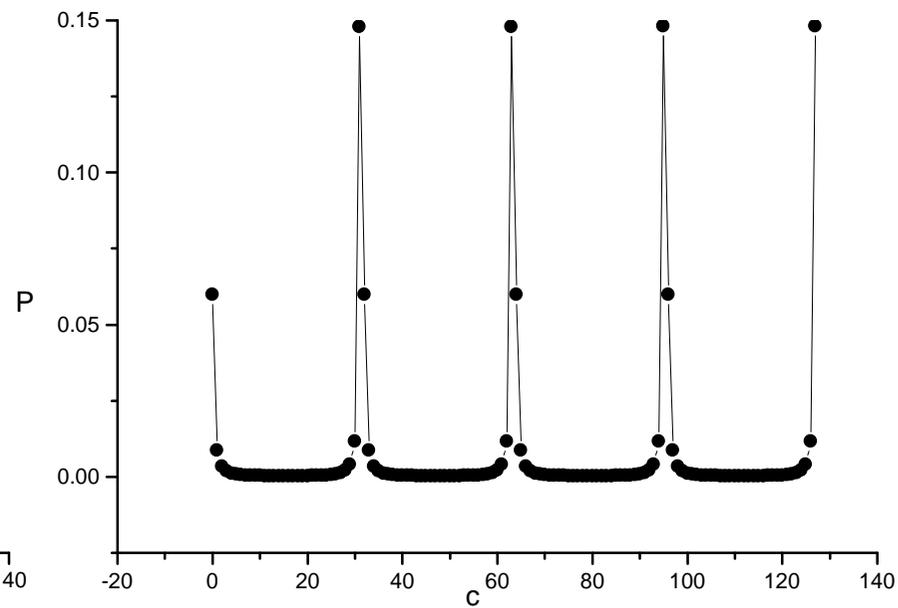
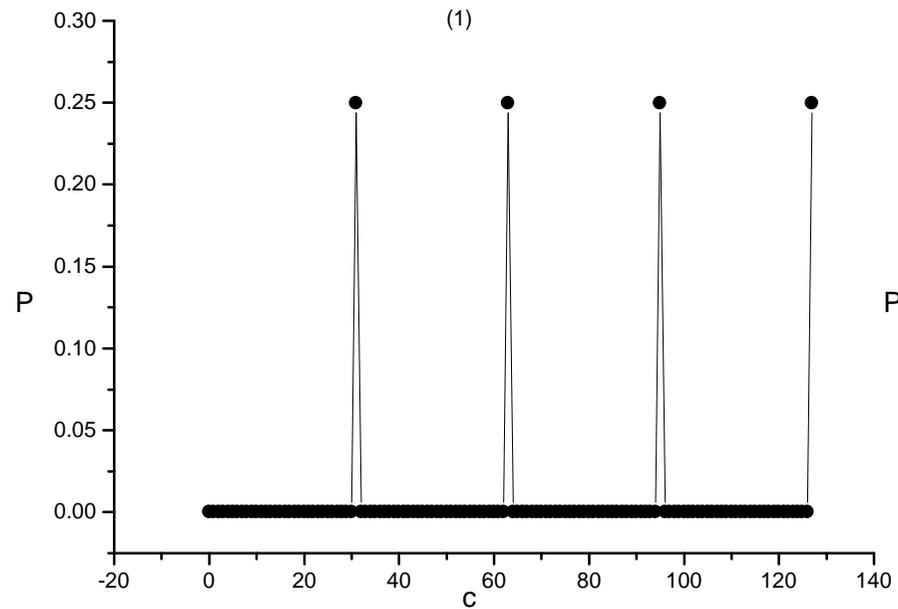
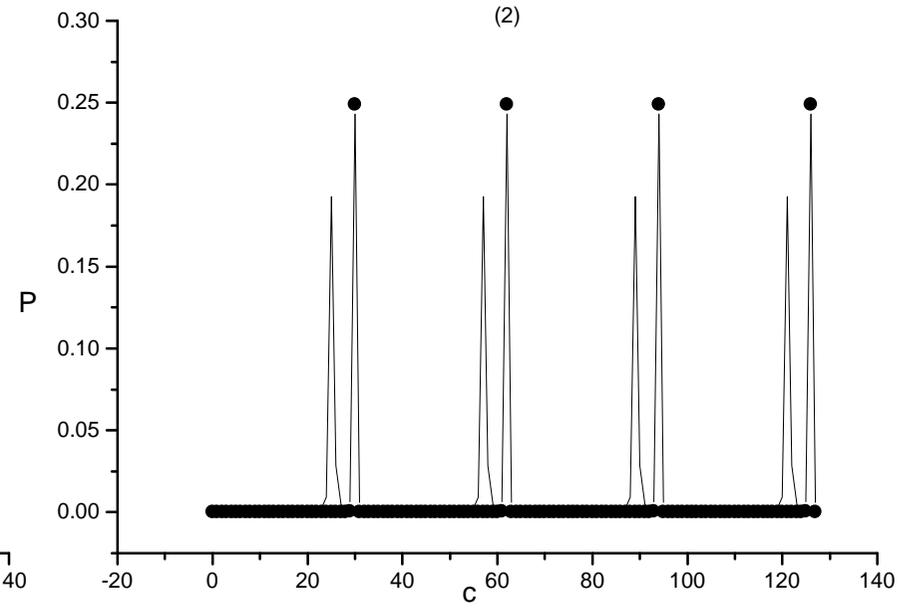
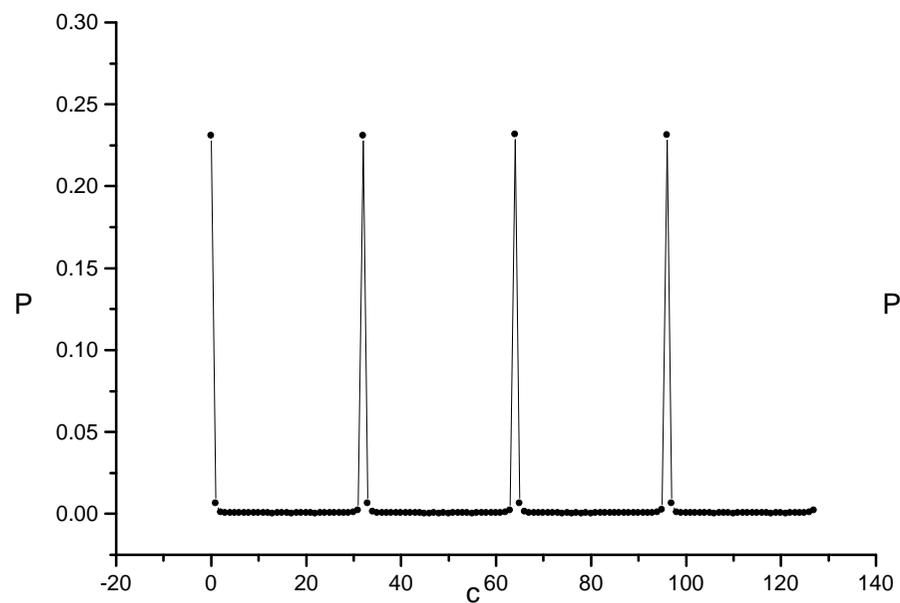
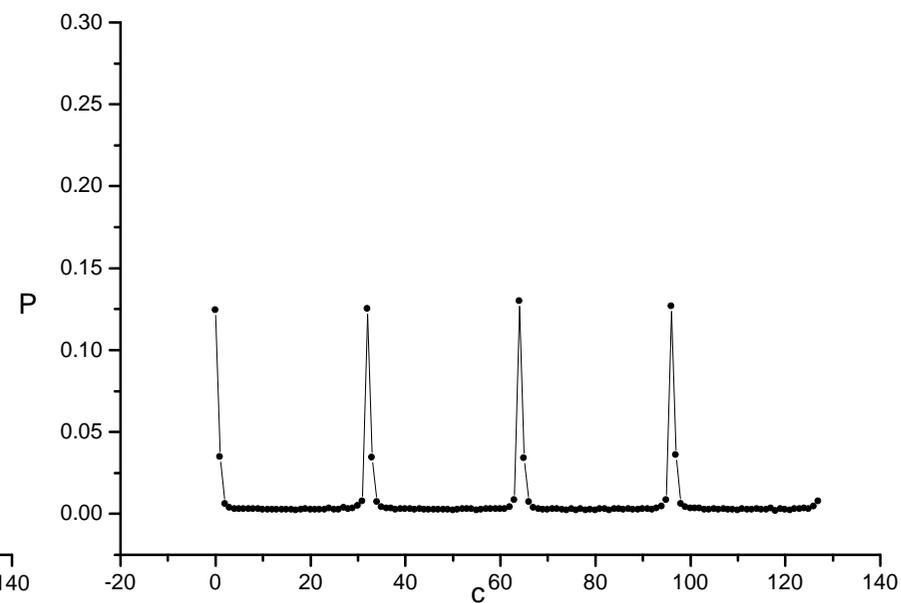
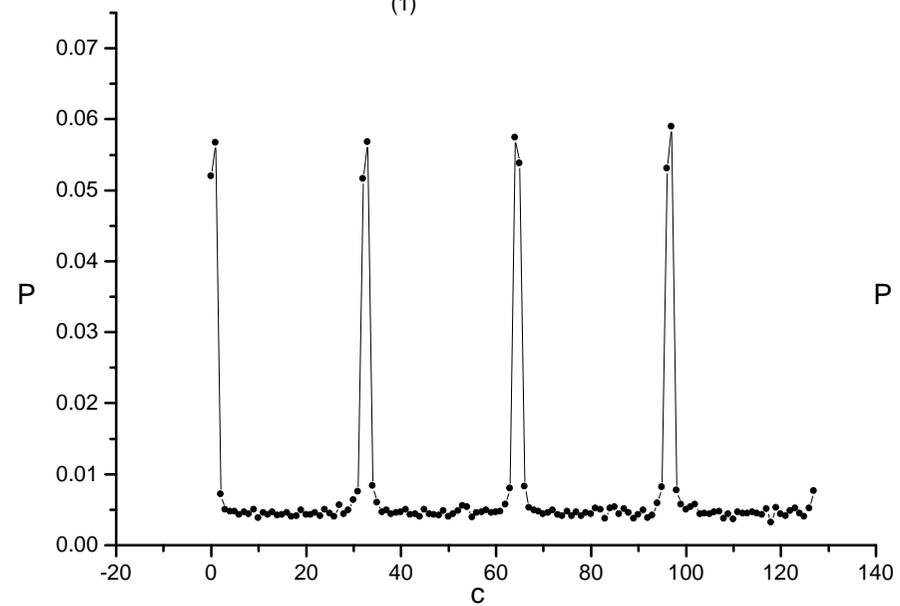
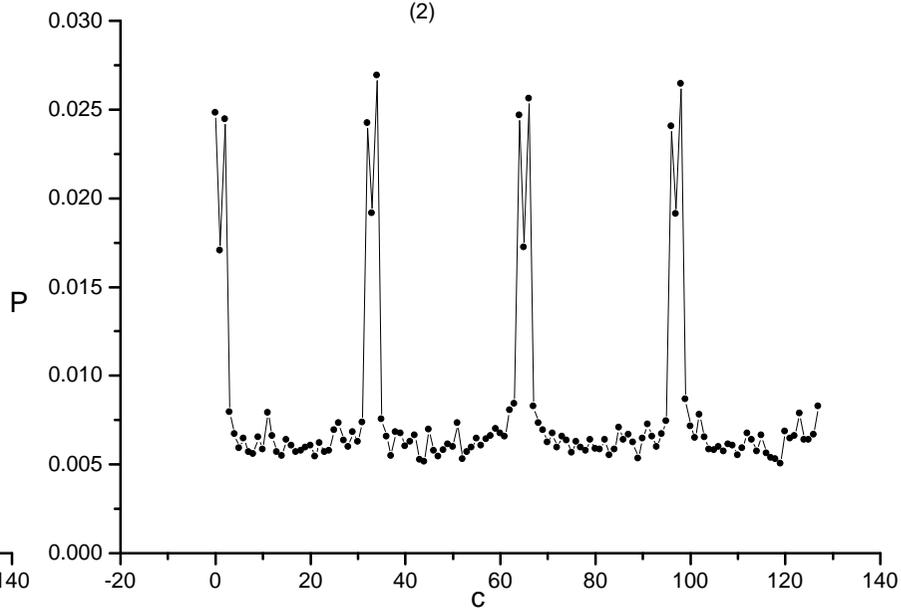
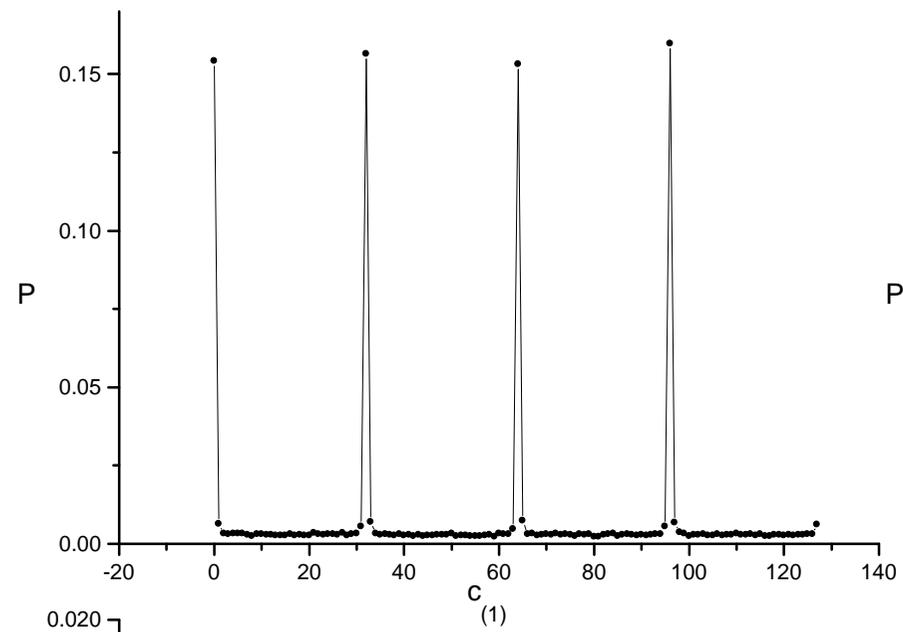
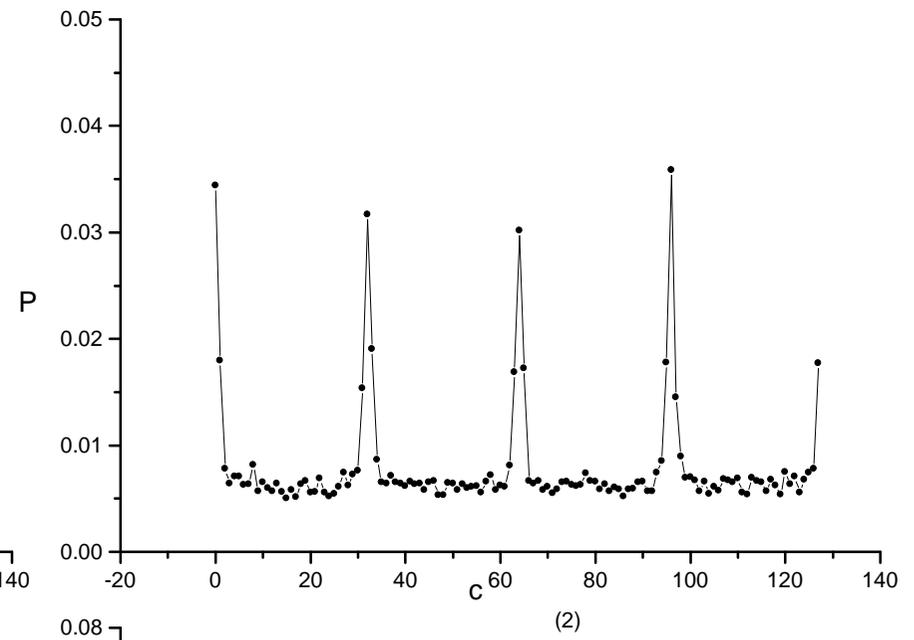
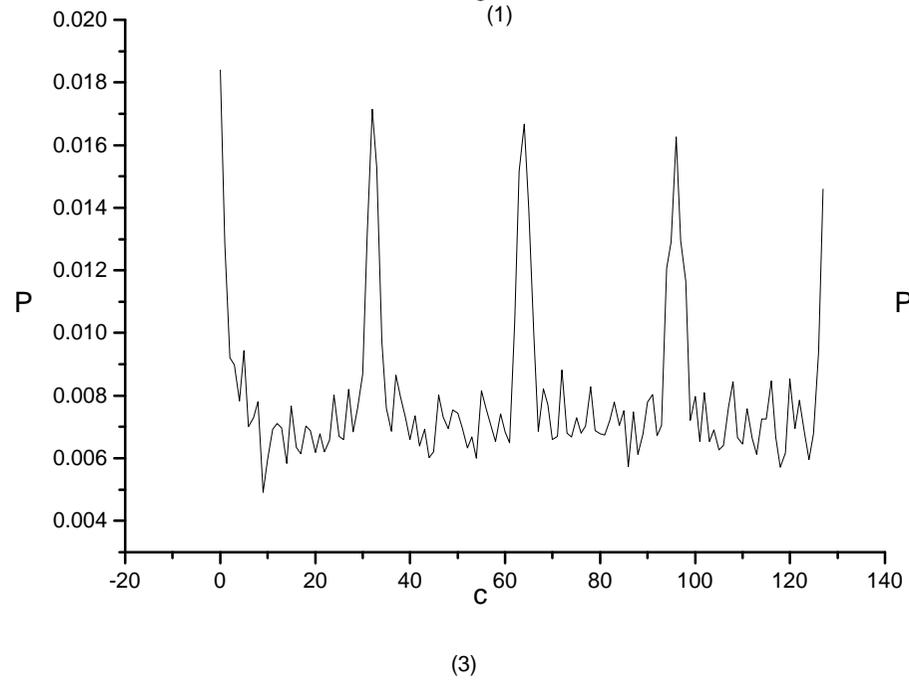
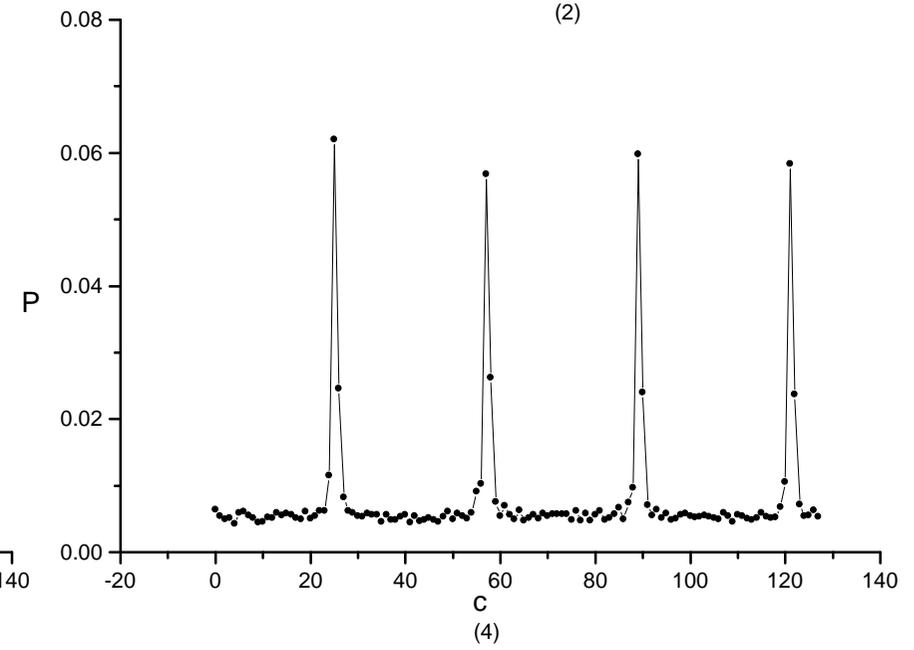
\end{document}